\documentclass[cameraready]{Interspeech}

\title{Collecting Prosody in the Wild: A Content-Controlled, Privacy-First Smartphone Protocol and Empirical Evaluation}

\author[affiliation={1,2}, orcid=0000-0001-6728-2063, correspondingauthor]{Timo K.}{Koch}
\author[affiliation={3,4}, orcid=0000-0002-5759-4976] {Florian}{Bemmann}
\author[affiliation={2,5}, orcid=0000-0001-7275-0626]{Ramona}{Schoedel}
\author[affiliation={2}, orcid=0000-0002-0597-8708]{Markus}{Buehner}
\author[affiliation={1}, orcid=0000-0002-4498-3067]{Clemens}{Stachl} 

\address{
    $^1$ Institute of Behavioral Science \& Technology, University of St. Gallen, Switzerland \\
    $^2$ Department of Psychology, LMU Munich, Germany \\
    $^3$ Social Data Science and AI Lab, LMU Munich, Germany \\
    $^4$ School of Social Sciences, University of Mannheim, Germany \\
    $^5$ Charlotte Fresenius Hochschule, University of Psychology, Germany
}

\email{timo.koch@unisg.ch}


\usepackage{comment}

\begin{document}

\maketitle

\begin{abstract}
Collecting everyday speech data for prosodic analysis is challenging due to the confounding of prosody and semantics, privacy constraints, and participant compliance. We introduce and empirically evaluate a content-controlled, privacy-first smartphone protocol that uses scripted read-aloud sentences to standardize lexical content (including prompt valence) while capturing naturalistic variation in prosodic delivery. The protocol performs on-device prosodic feature extraction, deletes raw audio immediately, and transmits only derived features for analysis. We deployed the protocol in a large study (N = 560; 9,877 recordings), evaluated compliance and data quality, and conducted diagnostic prediction tasks on the extracted features, predicting self-reported speaker sex and momentary affective states (valence, arousal). We discuss implications and directions for advancing and deploying the protocol.

\end{abstract}

\section{Introduction}

Speech research is increasingly moving from controlled laboratory settings to everyday life, for example, leveraging standard smartphones \cite{kaplanFablaVoicebasedEcological2025, khorramPRIORIEmotionDataset2018, birdAikumaMobileApp2014}. Although this shift enables large-scale and ecologically valid speech sampling, it also introduces two core challenges for prosodic data collection and analysis. First, real-world recordings are typically lexically unconstrained: what people say (semantics) varies as much as how they say it (prosody), for example, in the expression of affect \cite{kotzWhenEmotionalProsody2007, schwartzEmotionalSpeechProcessing2012}. As a consequence, semantics and prosody are confounded, making it difficult to isolate prosodic variation as an independent signal. Second, many in-the-wild approaches rely on collecting and storing raw audio. However, raw audio can contain speaker-informative cues and other potentially privacy-sensitive information, raising practical and legal hurdles (e.g., the General Data Protection Regulation) for large-scale deployment and reuse.

The prosody-semantics confound has long been addressed in laboratory research by instructing participants to read fixed lexical content through standardized passages and phonetically balanced sentence lists \cite{IEEERecommendedPractice1969}, and by using acted or read-speech corpora in which semantic content is controlled by design \cite{bussoIEMOCAPInteractiveEmotional2008}. However, comparable content-control techniques, and especially explicit control of lexical valence, that transfer the methodological rigor from the lab to the wild have rarely been integrated into smartphone protocols for speech data collection. Doing so could be particularly valuable given that a growing body of work leverages unconstrained speech, such as diary-style voice entries or passively captured snippets. In such settings, a scripted module could be added as a baseline to help separate individual-level prosodic variation from semantic content.

Moreover, much research on "in-the-wild" speech has relied on collecting and storing raw audio recordings \cite{weidmanNotHearingHappiness2020, khorramPRIORIEmotionDataset2018}. However, this practice raises significant privacy and ethical concerns because raw recordings may reveal speaker characteristics and contextual information beyond the intended research signal \cite{jessenForensicPhonetics2008,pohlhausenPrivacypreservingConversationAnalysis2026}. To mitigate these risks, recent work has shifted towards on-device processing where raw audio never leaves the user's device and prosodic features are extracted locally, creating a promising avenue for privacy-respectful research \cite{zhouOnDeviceSpeechFiltering2024, peplinskiFRILLNonSemanticSpeech2021}, but it also restricts validation and analysis choices, increasing the importance of careful study design and empirical analysis of the prosodic signal.

Here, we address the challenges of prosody-semantics confounds and privacy protection by introducing and empirically evaluating a content-controlled, privacy-first smartphone protocol for collecting prosodic speech data in the wild. The protocol standardizes speech content using validated, standardized read-aloud sentences, performs acoustic feature extraction on the device using openSMILE, immediately deletes raw audio, and transmits feature-level data only. We implemented this protocol in a large smartphone panel study (N = 560; 9,877 recordings) and evaluated participant compliance, data quality, and the characteristics of the resulting prosodic features. As validation, we use the extracted features to predict self-reported speaker sex and concurrently assessed momentary affect (valence, arousal). Our contribution is a reproducible, field-ready protocol for controlled, privacy-preserving speech data collection, along with evidence-based insights into its practical capabilities.

\section{Protocol}

\subsection{Voice Recording}

\begin{figure}[t]
  \centering
  \includegraphics[width=\linewidth]{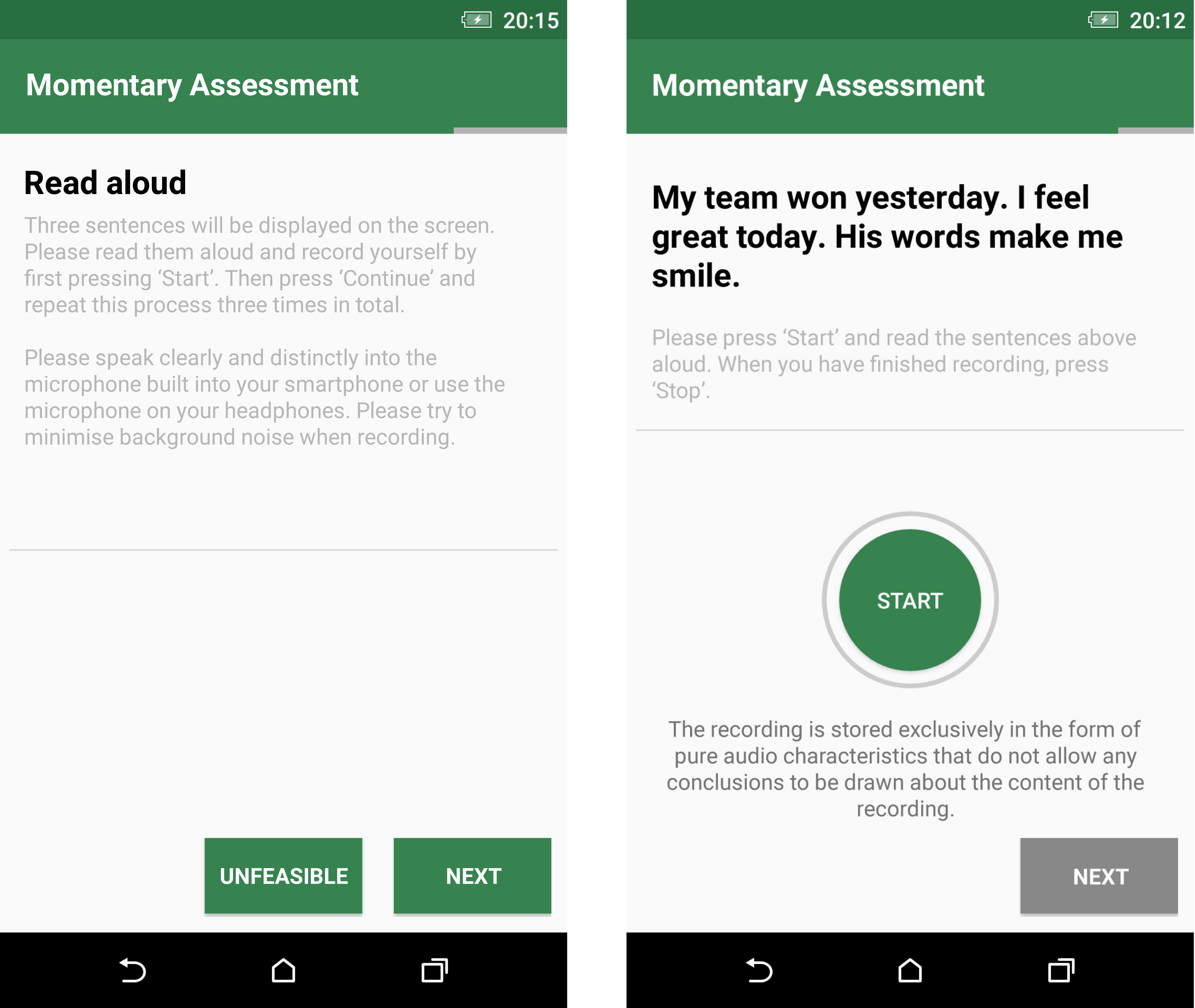}
  \caption{Graphical user interface of the smartphone-based voice recording (instructions translated to English). 
Left: Introductory screen with instructions and skip option.
Right: Recording screen in the “positive” lexical valence condition. }
  \label{fig:app_ui}
\end{figure}

The protocol was implemented as a module within the \textit{PhoneStudy} app's smartphone-based ecological momentary assessment (EMA) procedure, which prompted participants multiple times per day. At the start of each prompt, participants were shown a brief introductory screen describing the voice-recording task. Here, they also had the option to skip in case they were currently in an unfeasible situation (e.g., in a noisy location or in a situation where they could not talk). The second screen stated the three read-aloud sentences and an instruction on the screen on how to start and end the voice recording (see Figure~\ref{fig:app_ui}). The sentences presented to the participants were based on a set of 54 validated German sentences \cite{defrenEmotionalSpeechPerception2018}. They differed in their emotional valence: positive (e.g., “My team won yesterday.”), negative (e.g., “Nobody is interested in my life.”), and neutral (e.g., “The plate is on the round table.”). In each measurement instance, participants completed three separate speech recordings, corresponding to positive, neutral, and negative valence conditions, with the order of conditions randomly assigned per assessment. Within each recording, participants read three sentences of the respective valence, which were randomly drawn (with replacement) from the corresponding sentence set so participants would not get used to the content and know it by heart. We selected three sentences per recording to ensure sufficient vocal material (i.e., approximately 15 spoken words). Thus, each EMA yielded three recordings (one per valence condition). The audio recording was started by the participants via a button on the screen. Participants could stop the recording manually after a minimum of four seconds. Alternatively, the recording was stopped automatically after twelve seconds. We chose these lower and upper time thresholds because this was the minimum and maximum time required to read the three sentences per condition, determined by reading the sentences extremely fast and extremely slow and recording the times. 

\subsection{On-Device Prosodic Feature Extraction and Secure Data Transfer}

The on-device audio processing pipeline consisted of four sequential steps: local audio capture, on-device feature extraction, immediate deletion of raw audio, and secure batch transmission of feature data to the server. The voice recording user interface saved the audio of the read-aloud sentences in WAV audio file format. It contained uncompressed audio in the linear pulse-code modulation format (LPCM) at 16 bit depth and 44.1 kHz frequency. It avoided quality loss that could be imposed by compressed file formats, and required low processing power to have the app running smoothly. After the recording was completed, the feature extraction was started. We used the openSMILE open-source algorithm \cite{eybenOpensmileMunichVersatile2010} to automatically extract acoustic features directly on the participant’s device. Here, we used the extended Geneva Minimalistic Acoustic Parameters Set (eGeMAPS) which comprises 88 acoustic features \cite{eybenGenevaMinimalisticAcoustic2016} and the 2016 Interspeech Computational Paralinguistic Challenge (ComParE) set of 6,373 acoustic features \cite{schullerINTERSPEECH2016Computational2016}. The openSMILE algorithm was embedded as an executable file, specifically built for the Android ARM architecture as an Android native library. It ran through a shell command that passed the local path to the audio recording, the target path for the resulting features, extraction configuration options, and the openSMILE configuration files (eGeMAPSv01a.conf and ComParE\_2016.conf). The resulting features were then stored as a CSV file, and were immediately transferred into the research app’s data architecture. All local files, including the WAV audio file and the resulting CSV file, were deleted afterwards. Data was then temporarily stored in a local, in-app database which was private to the app, i.e. could not be accessed by other apps or the user. In regular intervals at opportune moments (i.e., when the smartphone was not used and had WiFi connection) the app synchronized its local data with the remote server. Then, the feature vectors were transmitted. Data were transferred through an SSL encrypted transmission, and a three-way handshake was used to reduce risk of transfer failures. After each transmission, the local data were deleted from the participant’s phone. This increased security and privacy and reduced the consumption of storage space in the app.

\section{Evaluation}

\subsection{Data Collection}

The protocol was implemented in a data collection that was part of a large panel study \cite{schoedelBasicProtocolSmartphone2024}. Data collection was approved by the ethics committee of the psychology department at LMU Munich and all procedures adhered to the General Data Protection Regulation (GDPR). We recruited a quota-matched sample of N = 850 participants representative of the German population in terms of age, sex, education, income, religious affiliation, and relationship status. Participants had to be between 18 and 65 years old, fluent in German, and possess an Android smartphone (Android 5 or higher) for private use as a sole user. Informed consent was obtained from all participants included in the study. The study comprised two two-week ecological momentary assessment (EMA) phases (July 27, 2020, to August 9, 2020; September 21, 2020, to October 4, 2020) during which participants were prompted to complete two to four EMAs per day on their personal smartphone. The last EMA prompt of each day included our voice data collection protocol. We chose to include it then to increase participation because we assumed it would be higher in the evening, when people are more likely to be alone and at home. The audio recording occurred at the end of the EMA instance and the participants had the option to skip. With this approach, we collected 11,217 recordings from 3,813 EMA instances from 578 participants in total. Participants made on average 19.41 (SD = 12.37) voice recordings. 

\subsection{Participant Compliance and Feature-Based Filtering}

Overall, compliance with the voice protocol was good. Participants initiated the recording sequence in 67.8\% of the prompts (3,813 of 5,627) and selected “skip” in 1,814 prompts, which compares favorably with related work collecting scripted speech via smartphones \cite{dineleyRemoteSmartphoneBasedSpeech2021}. Conditional on initiation, participants completed all three recordings (pos/neu/neg) once started in 96.9\% of instances.

Given the scale and privacy-preserving design of the study, we used feature-based filters to exclude likely invalid speech recordings. First, we used openSMILE-derived descriptors (including voicing probability and segment statistics) to flag recordings unlikely to contain human speech. Specifically, we flagged human voice as absent for a recording if any of the following held: mean voicing probability $< 0.5$, voiced segments per second $= 0$, or mean voiced-segment length $= 0$. Thus, we dropped 232 voice records (from 46 participants), likely not containing human speech. As an additional robustness filter, we excluded 1,108 clips (from 235 participants) with non-positive harmonic-to-noise ratio (HNR $\le 0$ dB), indicating that the aperiodic component is at least as strong as the periodic (harmonic) component in voiced frames \cite{Boersma1993HNR}. This left us with a final set of prosodic features from 9,877 voice samples from 3,513 EMA instances from 560 participants (46\% female; Age$_M$ = 41.80 years, Age$_{\text{SD}}$ = 12.80 years). Thus, there were on average 17.64 (SD $= 11.70$) voice records per participant. 

\subsection{Acoustic Properties of Collected Speech}

\begin{figure}[t]
  \centering
  \includegraphics[width=\linewidth]{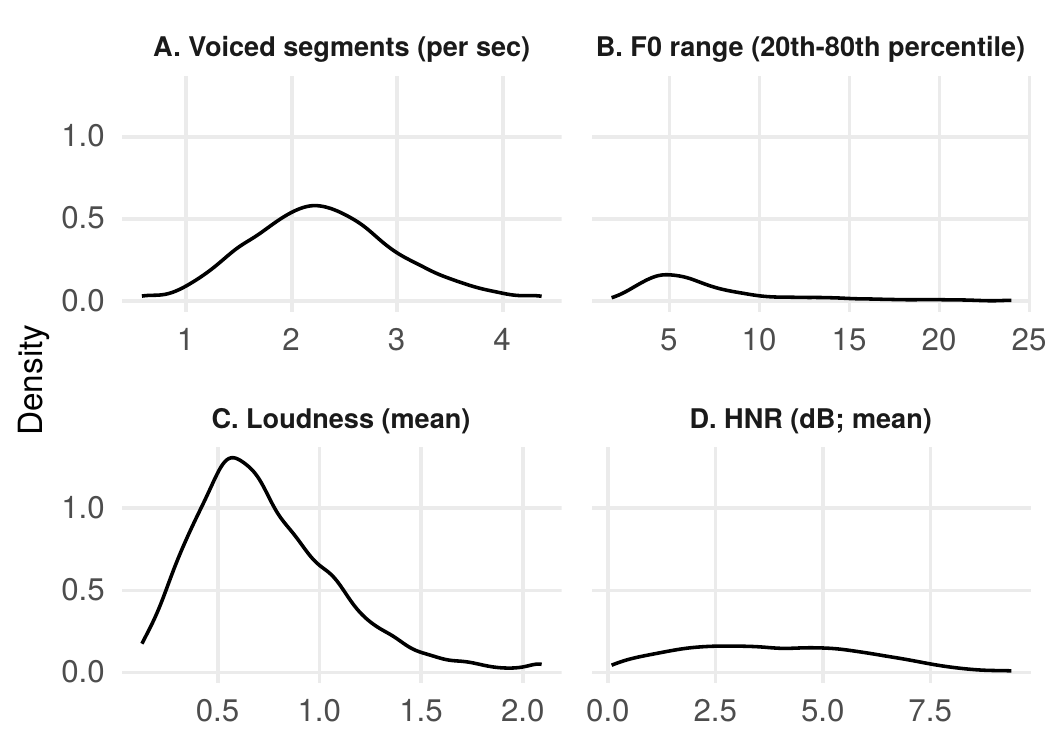}
  \caption{Distributions of selected key acoustic diagnostics across recordings.}
  \label{fig:acoustic_diagnostics}
\end{figure}

To assess whether the proposed protocol yielded analyzable acoustic speech in naturalistic smartphone settings, we examined the distributional properties of key acoustic diagnostics of recordings retained after filtering (see Figure~\ref{fig:acoustic_diagnostics}). Across recordings, voiced segments per second were concentrated away from zero, indicating that retained clips contained voiced speech. F0 percentile-range variability was non-zero, indicating measurable prosodic variation in the retained data. Loudness exhibited a broad distribution, reflecting variation in speaker vocal intensity and recording conditions. Finally, HNR values were concentrated in the positive range, consistent with the exclusion of low-signal and highly noisy recordings without collapsing variability. Together, these patterns suggest that the retained speech samples collected via this protocol contain analyzable prosodic information while preserving the variability characteristic of in-the-wild recordings.

To further characterize measurement properties of the collected prosodic features, we tested for systematic differences across prompt-set conditions and quantified within-speaker stability. For each of the four key prosodic metrics in Figure~\ref{fig:acoustic_diagnostics}, we fit linear mixed-effects models with condition as a fixed effect and a random intercept for participant. Outcomes were $z$-standardized across recordings to facilitate comparison across metrics. Condition effects were evaluated via contrasts against the neutral condition; within-speaker stability was summarized via participant intraclass correlations (ICC$_p$) from the same variance decomposition. To account for multiple testing across the four metrics and two contrasts, we report Benjamini--Hochberg FDR-corrected $p$ values (Table~\ref{tab:prosody_mlm}).

Condition effects were small overall. F0 variability did not differ reliably across conditions (both FDR-adjusted $p \ge .348$). In contrast, mean HNR and voiced segments per second showed modest but consistent shifts (both $|\beta| \approx 0.06$--$0.13$ SD; all $p < .001$), and mean loudness was slightly lower in both positive and negative prompts ($\beta=-0.036$ and $-0.034$ SD; $p=.020$ and $.025$). Speaker-level stability was substantial: between-participant differences accounted for 32.5\%--69.3\% of variance across metrics (ICC$_p$ range: 0.325--0.693). These results indicate that the protocol yields speaker-informative prosodic summaries with only modest prompt-set-related shifts.

\begin{table}[t]
\centering
\caption{Condition contrasts (vs.\ neutral) from linear mixed-effects models for four prosodic metrics, with a random intercept for participant. Outcomes were $z$-standardized across recordings. Reported $p$ values are Benjamini--Hochberg FDR corrected across the 8 tests in the table. ICC$_p$ denotes the intraclass correlation attributable to participant (variance proportion).}
\label{tab:prosody_mlm}
\scriptsize
\setlength{\tabcolsep}{2pt}
\resizebox{\columnwidth}{!}{%
\begin{tabular}{llrrrrr}
\toprule
Metric & Contrast & $\beta$ & SE & $t$ & $p$ & ICC$_p$ \\
\midrule
F0 range (20--80 perc)      & Neg--Neu &  \phantom{-}0.021 & 0.020 &  \phantom{-}1.03 & .348 & 0.350 \\
F0 range (20--80 perc)      & Pos--Neu & -0.005           & 0.020 & -0.22            & .824 & 0.350 \\
HNR (mean)               & Neg--Neu & -0.071           & 0.014 & -5.03            & \textless .001 & 0.693 \\
HNR (mean)               & Pos--Neu &  \phantom{-}0.060 & 0.014 &  \phantom{-}4.26 & \textless .001 & 0.693 \\
Voiced seg/s             & Neg--Neu &  \phantom{-}0.107 & 0.021 &  \phantom{-}5.24 & \textless .001 & 0.325 \\
Voiced seg/s             & Pos--Neu & -0.128           & 0.020 & -6.27            & \textless .001 & 0.325 \\
Loudness (mean)          & Neg--Neu & -0.034           & 0.015 & -2.35            & .025 & 0.673 \\
Loudness (mean)          & Pos--Neu & -0.036           & 0.015 & -2.49            & .020 & 0.673 \\
\bottomrule
\end{tabular}%
}
\end{table}

\subsection{Protocol Validation via Prediction Tasks}

We used two downstream prediction tasks as diagnostic checks of whether the protocol produced speaker- and state-informative acoustic measurements. First, we assessed whether the extracted features retained stable speaker-related information using a basic prediction task: predicting participants’ self-reported sex from prosodic features. We trained a random forest classifier ($\texttt{num.trees}=1000$, $mtry=\lfloor\sqrt{p}\rfloor$, $\texttt{min.node.size}=1$) on extracted features with sex (male, female) as the target, using ten-fold cross-validation with participant-level blocking (i.e., all recordings from a participant were assigned exclusively to either the training or test set within each fold). Performance was equally high for the eGeMAPS (balanced accuracy$_{\mathrm{Md}}$ = 91.77\%; SD across CV folds  = 3.11\%) and ComParE (balanced accuracy$_{\mathrm{Md}}$ = 92.04\%; SD = 3.12\%) feature sets.

We then examined whether prosodic features were predictive of concurrently self-reported affective states in a preregistered analysis \cite{kochPredictingAffectiveStates2021}. Momentary valence and arousal were assessed with a single item on a 6-point Likert scale \cite{russellCircumplexModelAffect1980}. Valence was assessed with “How do you feel right now?” (1 = very unpleasant, 6 = very pleasant) and arousal with “How would you rate your current level of activation?” (1 = very inactive, 6 = very activated). For each outcome, we trained a random forest regressor ($\texttt{num.trees}=1000$, $mtry=\lfloor\sqrt{p}\rfloor$, $\texttt{min.node.size}=5$) to predict affect ratings and evaluated out-of-sample performance using participant-blocked ten-fold cross-validation. Prediction performance for momentary affect was modest. Using eGeMAPS features, performance was $\rho_{Md}=0.12$ ($\rho_{SD}=0.06$) for arousal and $\rho_{Md}=0.02$ ($\rho_{SD}=0.08$) for valence. Using the larger ComParE feature set yielded only slightly higher performance for arousal ($\rho_{Md}=0.13$; $\rho_{SD}=0.07$) and no improvement for valence ($\rho_{Md}=0.03$; $\rho_{SD}=0.06$), suggesting that limited performance was not primarily driven by feature-set size. To assess whether prediction error differed across sentence-valence conditions, we repeated the prediction analyses separately for each condition (positive, neutral, negative). We compared fold-wise MAE across sentence-valence conditions using a Friedman test and found no evidence for condition differences for either outcome (both $p>0.05$).

\subsection{Code and Data Availability}

Analysis scripts and aggregated results for the empirical evaluation are available at \url{https://github.com/Timo-Ko/prosody_in_the_wild}. We additionally provide a demonstration implementation of the on-device audio processing pipeline as an Android module at \url{https://github.com/Flo890/demo-prosody-in-the-wild}. 

\section{Discussion}

\subsection{Protocol Contributions}

The primary contribution of this work is a field-ready protocol for collecting prosodic speech data in everyday life that jointly addresses two  challenges in naturalistic speech research: confounding between semantic content and prosody and the privacy challenges associated with raw audio collection. The protocol standardizes lexical content using validated lexically valence-balanced read-aloud sentences, performs acoustic feature extraction on device, and deletes raw audio immediately, allowing repeated prosodic measurement at scale. The empirical evaluation demonstrates that this approach is feasible in practice and yields analyzable, feature-level speech data with good participant compliance under realistic conditions.  

Moreover, the prediction analyses provide an empirical characterization of the signal available under this protocol: While speaker characterization (i.e., sex classification) yielded strong results, the prediction of affect self-reports was weak overall, with slightly stronger signal for arousal than for valence and no reliable differences across sentence-valence conditions. This is consistent with previous work using unconstrained daily-life voice samples, which similarly reports weak predictive performance for affective states from prosodic descriptors \cite{weidmanNotHearingHappiness2020}, and with reports that arousal tends to be more reliably associated with acoustic variation than valence \cite{banseAcousticProfilesVocal1996}. Several factors may contribute to the limited observed performance, including limited affective expression in scripted read-aloud speech and heterogeneous recording conditions in naturalistic smartphone use, such as device placement, background noise, and microphone characteristics, which can dilute prosodic signal and reduce generalization \cite{busquetVoiceAnalyticsWild2024}. In addition, the use of single-item EMAs, while common practice \cite{dejonckheereAssessingReliabilitySingleitem2022}, may introduce measurement error that further limits attainable inference. However, the protocol is not limited to using extracted prosodic features as predictors. It can also serve as a methodological baseline alongside naturalistic voice collection, providing within-participant reference measurements of typical prosodic range and day-to-day variability that can be used to calibrate analyses of unconstrained speech and control for unsystematic vocal fluctuations in real-world recordings.

Finally, the privacy-first design addresses key obstacles inherent in everyday audio recording. Because participants read pre-defined prompts, they are not required to disclose personal or sensitive content to bystanders, and on-device processing ensures that raw voice recordings never leave the participant’s device. This design follows a precautionary data-minimization rationale: forensic-phonetic work treats speech as potentially informative for speaker profiling, while also emphasizing that such evidence must be interpreted in context rather than as deterministic identification \cite{jessenForensicPhonetics2008}.  At the same time, residual privacy risks remain: feature representations can be difficult to communicate to participants in terms of their informational content, and it is not fully understood which higher-level personal attributes may be inferable from such features. Consequently, while the protocol substantially reduces privacy risks relative to raw audio collection, it cannot eliminate the possibility that sensitive characteristics may be inferred with future modeling advances \cite{tomashenkoVoicePrivacy2020Challenge2022}.

\subsection{Limitations and Future Directions}

The findings should be interpreted in light of two main limitations. First, although the instructions prompted participants to read the sentences displayed verbatim, compliance could not be guaranteed. We implemented feature-based filtering (e.g., human-voice detection), but we did not retain raw audio that would allow verification of lexical fidelity post hoc. As a result, occasional deviations (e.g., paraphrasing, disfluencies) could have reintroduced semantic variance and altered prosody. Future work could incorporate lightweight on-device checks, such as ASR-based keyword matching without storing raw text or audio, to flag non-reads and improve quality control \cite{sainathConvolutionalNeuralNetworks2015}. In addition, dedicated calibration studies using parallel lab-recorded and smartphone-recorded speech could help establish how feature distributions and quality indicators derived from privacy-preserving protocols relate to manually verified reference recordings.
Second, the protocol relies on engineered acoustic features that favor comparability, interpretability, and efficient on-device processing over maximal representational capacity. Contemporary self-supervised audio embeddings may capture additional acoustic structure and improve performance in downstream prediction tasks. Future work could therefore evaluate lightweight embedding models that can be computed on the device  \cite{peplinskiFRILLNonSemanticSpeech2021} within the same privacy-preserving protocol. 

\section{Conclusion}

We presented a content-controlled, privacy-first smartphone protocol for collecting prosodic speech data in everyday life and empirically evaluated it. By standardizing lexical content, performing on-device acoustic feature extraction, and deleting raw audio immediately, the protocol enables scalable, privacy-preserving prosodic data collection in naturalistic settings.

\newpage
\section{Acknowledgments}

We thank audEERING GmbH for their support with integrating openSMILE into the smartphone app, and Peter Ehrich and Dominik Heinrich for their support with the technical implementation of the on-device voice feature extraction. We also thank the Leibniz Institute for Psychology (ZPID) for funding data collection. This project was supported by the Swiss National Science Foundation (SNSF) under project number 215303 and a scholarship of the German Academic Scholarship Foundation.

\section{Generative AI Use Disclosure}
We used ChatGPT (OpenAI) to improve code scripts and copy editing in the form of small improvements to human-generated text for readability and style. All outputs were reviewed and edited by the authors, who assume full responsibility for the content.

%
%

\bibliographystyle{IEEEtran}
\bibliography{prosody_in_the_wild}

\end{document}